\documentclass[ijfs,article,accept,pdftex,moreauthors]{Definitions/mdpi} 
\usepackage[acronym,nohypertypes={acronym,notation}]{glossaries}
\newacronym{lop}{LOP}{Law of one price}
\newacronym{ltcm}{LTCM}{Long-Term Capital Management L.P.}
\newacronym{tw}{TW}{Trading Weights}
\newacronym{ssd}{SSD}{Sum of Squared Deviation}
\newacronym{dm}{DM}{Distance Method}
\newacronym{ott}{OTT}{Optimal Trading Technique}
\usepackage{makecell}
\usepackage{algorithmic}
\usepackage[lined,boxed,ruled,commentsnumbered]{algorithm2e}
\firstpage{1} 
\pubvolume{1}
\issuenum{1}
\articlenumber{0}
\pubyear{2024}
\copyrightyear{2024}
\externaleditor{Academic Editor: Babacar Sène}
\datereceived{17 June 2024} 
\daterevised{2 August 2024} 
\dateaccepted{5 August 2024} 
\datepublished{ } 
\hreflink{https://doi.org/} 
\Title{Optimal Market-Neutral Multivariate Pair Trading on the Cryptocurrency Platform}

\TitleCitation{Optimal Market-Neutral Multivariate Pair Trading on the Cryptocurrency Platform}

\Author{Hongshen Yang\orcidA{} * and Avinash Malik\orcidB{}} 

\AuthorNames{Hongshen Yang and Avinash Malik}

\AuthorCitation{Yang, Hongshen, and Avinash Malik}

\address[1]{%
Department of ECSE, Faculty of Engineering, The University of Auckland, Auckland CBD, \linebreak Auckland 1010,
 New Zealand; avinash.malik@auckland.ac.nz}

\corres{\hangafter=1 \hangindent=1.05em \hspace{-0.82em} Correspondence: hyan212@aucklanduni.ac.nz}

\abstract{This research proposes a novel arbitrage approach in multivariate pair trading, termed the Optimal Trading Technique (OTT). We present a method for selectively forming a ``bucket'' of fiat currencies anchored to cryptocurrency for monitoring and exploiting trading opportunities simultaneously. To address quantitative conflicts from multiple trading signals, a novel bi-objective convex optimization formulation is designed to balance investor preferences between profitability and risk tolerance. We understand that cryptocurrencies carry significant financial risks. Therefore this process includes tunable parameters such as volatility penalties and action thresholds. In experiments conducted in the cryptocurrency market from 2020 to 2022, which encompassed a vigorous bull run followed by a bear run, the OTT achieved an annualized profit of 15.49\%. Additionally, supplementary experiments detailed in the appendix extend the applicability of OTT to other major cryptocurrencies in the post-COVID period, validating the model's robustness and effectiveness in various market conditions. The arbitrage operation offers a new perspective on trading, without requiring external shorting or holding the intermediate during the arbitrage period. As a note of caution, this study acknowledges the high-risk nature of cryptocurrency investments, which can be subject to significant volatility and potential loss.}


\begin{document}

\section{Introduction}
\label{sec:introduction}

The secondary market functions as a modern, efficient form of double auction where participants intending to trade assets submit their acceptable bid and ask prices. The highest bid and lowest ask are matched if the prices meet. We believe in the Efficient Market Hypothesis (EMH), which posits that in the secondary market, prices should reflect all available information \citep{fama_efficient_1970}. However, historical economic bubbles show that investors do not always act rationally. \cite{dale_financial_2005} illustrated how an order book filled with irrational bids and asks could drive prices in the wrong direction. Identifying inefficiencies in an efficient market helps maintain trading equilibrium. Market directional trading involves astute traders estimating more accurate prices for instruments. They then place orders with the expectation that the market will revert to rationality. Traders who bet correctly are rewarded by the market mechanism, contributing to price discovery. Quantitative trading, instead of relying on experienced accountants or financial analysts, utilizes mathematical modeling for better price estimations.

The cryptocurrency market is an emerging and yet immature secondary market. The recent cryptocurrency market trend benefits from technological advancements and global participation. Many people criticize the significant investment risk; however, its volatility, liquidity, and 24/7 trading make it an ideal platform for algorithmic trading. The cryptocurrency exchange users can trade between cryptocurrencies and fiat currencies, exploiting inefficiencies and capturing anomalies from both forex \mbox{and cryptocurrency volatility. }

Our research focuses on pair trading, a widespread strategy based on the Law of One Price (\acrshort{lop}), suggesting that homogeneous items should have the same price everywhere. However, real world restrictions like trade frictions and legal regulations often violate this principle \citep{isard_how_1977, miljkovic_law_1999}. Similar assets should have similar price trends. However, their price independence can cause divergence, and convergence can often result in profit. Arbitrage-based trading approaches, grounded in \acrshort{lop}, are perceived as having more stable profitability. Investment banks and hedge funds have used pair trading for decades \citep{gatev_pairs_2006, elliott_pairs_2005, burgess_computational_2000}. While pair trading is common, the extension of pair trading to many-pair (multivariate) trading is not well studied. This paper delves into pair trading and introduces an optimization-based technique for multivariate assets, tested in the cryptocurrency market.

This research has several novel contributions:
\textcircled{1}~Unlike classical techniques \citep{perlin_m_2007, galenko_trading_2012, dunis_cointegration_2005}, our approach uses optimization to arbitrage multiple pairs simultaneously to the individual asset level without grouping assets.
\textcircled{2}~The bi-objective optimization is adjustable to investor preferences, including tunable parameters like risk tolerance and operating thresholds for maximizing return and minimizing risk.
\textcircled{3}~Our arbitrage approach avoids holding volatile intermediate cryptocurrency, maintaining positions in fiat currencies to lower risk.
\textcircled{4}~No borrowing is required for shorting, as all trading actions are achievable within the initial principal.
\textcircled{5}~The proposed technique is robust, market-neutral, and profitable in both bull and bear markets.

The remainder of the article is structured as follows. Sections~\ref{sec:relatedwork} review the most cited works in pair trading and outline the proposed trading idea. Sections~\ref{sec:opti-cont} introduce the prerequisite knowledge and details of the proposed trading methodology, respectively. Section~\ref{sec:simulation} presents the simulation algorithm. Finally, Sections~\ref{sec:results} and~\ref{sec:conclusions} present the results and conclusions, respectively. Additional experiments and findings are detailed in Appendix \ref{sec:sensitivity}.

\section{Related Work}
\label{sec:relatedwork}

\subsection{Pair Trading}
\label{sec:pairtrade}

\acrfull{ltcm} was a highly leveraged hedge fund founded by John Meriwether in 1994 that operated on bond pair trading. It is famous for its billions of dollars bailout after the 1997 financial crisis and is known as an early real-life example of pair trading \citep{slivinski_too_2009}.

\begin{figure}[H]
  
 \hspace{-20pt} \includegraphics[width=0.8\textwidth]{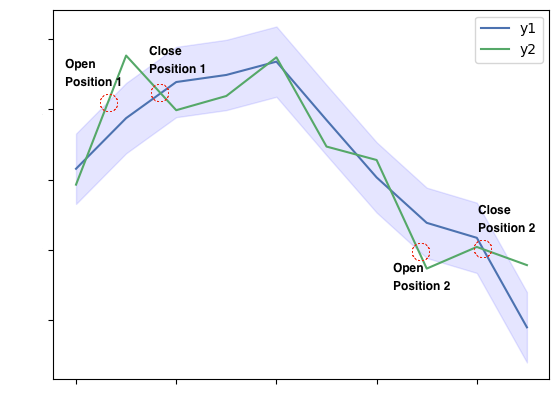}
  \caption{Pair
 trading between assets $y_1$ and $y_2$ with threshold band.}
  \label{fig:pairtrading}
\end{figure}

Pair trading is straightforward, as shown in Figure~\ref{fig:pairtrading}, which illustrates the most cited baseline framework---the Distance Method \citep{gatev_pairs_2006}. The common trading terminology includes \textit{long} positions as buying assets to sell later at a higher price, and \textit{short} as borrowing and selling assets, intending to buy them back at a lower price. We can summarize pair trading in the following steps:

\begin{enumerate}
  \item Search for two assets with the closest historical price movements, $y_1$ and $y_2$, and group $y_1$ and $y_2$ as a \textit{trading pair}.
  \item Calculate the price difference between $y_1$ and $y_2$ as \textit{distance}. 
  \item Open a position when the distance surpasses two times its historical standard deviation (\textit{open threshold}, shown as the band area with $y_1$) by shorting $y_2$ and longing $y_1$.
  \item Close the position when $y_1$ and $y_2$ prices converge (\textit{close threshold}), then close the position by reversing the buy and sell actions.
  \item If the price of $y_2$ falls below the threshold, open a position by doing the opposite: long $y_2$ and short $y_1$ until the next crossing point. 
  \item Repeat this practice until the end of the trading period.
\end{enumerate}

A critical property of pair trading is achieving \textit{market neutrality} by betting equivalent amounts on the long and short positions. Since the loss on one position is compensated by the other position, the portfolio value tends to remain stable with the overall market direction. \cite{chen_pricing_2018} formed pairs from asset returns instead of prices, suggesting that a temporarily weaker return implies a stronger rebound and vice versa. Both \cite{gatev_pairs_2006} and \cite{chen_pricing_2018} based their work on correlation.

Besides correlation, cointegration is a key concept in time-series analysis. Cointegration represents a long-run relationship between two time-series data sets \citep{engle_long-run_1991}. \cite{vidyamurthy_pairs_2004} examined assets for cointegration using the Engle--Granger test and combined assets based on correlation and cointegration, providing guidance for cointegration-based pair trading.

Further studies discuss multivariate pair trading, hedging market risk by managing positions and creating artificial pairs from scenarios with more than two assets. \cite{perlin_m_2007} developed a technique combining assets into an artificial index to form a pair. Instead of finding a pair, he created a synthetic index $P_{it}^*$ to anchor a real index $P_{it}$. The synthetic index $P_{it}^*$ was constructed by weighting ($w$) $m$ assets based on their prices:
\[P^*_{it} = \sum_{k=1}^m w_k P_{kt}^*\]

Weights among the participating assets were tuned to match $P_{it}$ as closely as possible. Next, $P_{it}$ and $P_{it}^*$ were arbitraged for pair trading. Such an artificially formed pair-trading methodology sometimes also uses cointegration techniques. \cite{galenko_trading_2012} used a moving window method to pair trade among a group of assets, calculating the cointegration vector $\boldsymbol{b}$ to track $Z_t$ from the asset return combined with the cointegration value $b$:
\[Z_t = \sum_{i=1}^N b^i r_t^i\]

Assets were then partitioned into two categories based on \highlighting{$\boldsymbol{b}$}
results: $L$ (long) and $S$ (short). Quantified $Z_t$ provided recommendations on investment amounts for each asset. \cite{dunis_cointegration_2005} adopted a similar methodology, tracking the EURO STOXX50 and split outperformers and underperformers into STOXX50+ and STOXX50-. They pair traded STOXX50+ and STOXX50- to achieve ``double alpha'', a weak form of pair trading due to non-negligible market risk.

There are many practices for pair trading and even multivariate pair trading techniques, but most existing pair trading techniques focus on compounding pairs rather than focusing on the individual granularity. Our research aims to provide a technique that neither creates an artificial index nor binds a group of assets. Using optimization techniques, we can operate directly on all participating assets to provide an optimal solution for multivariate pair trading.

\subsection{Cryptocurrency}
\label{sec:cryptocurrency}

The work in~\cite{fang_cryptocurrency_2022} provides a comprehensive survey of the history and recent research on cryptocurrencies. The cryptocurrency began with ~\cite{nakamoto_bitcoin_2008}'s groundbreaking work in 2008, introducing a decentralized monetary system based on a peer-to-peer ledger recording system. The token that operates within the monetary system is called Bitcoin (BTC). By incentivizing Bitcoin users to record a full transaction ledger on their own computers, the ledgers are distributed across countless individual nodes instead of a centralized server. This property enables frictionless international participation in Bitcoin. Furthermore, international participation provides excellent liquidity for Bitcoin trades. Shortly after, Bitcoin's success led to the creation of imitators. Ethereum (ETH), the second most famous cryptocurrency, was created by \cite{buterin_next-generation_2014} for its programmable extensibility. ETH gained its reputation from its fast development and widespread adoption in recent years \citep{bai_evolution_2020}. 

Popular pair trading techniques from Section~\ref{sec:pairtrade} have also been introduced in the cryptocurrency market. \cite{fil_pairs_2020} applied the distance method based on the work of~\cite{gatev_pairs_2006}, achieving a 3\% monthly profit in the cryptocurrency market. \cite{broek_cointegration-based_2018} formed pairs based on 60-day trading periods to profit from arbitrage opportunities, demonstrating existing inefficiencies of the cryptocurrency market.

\begin{figure}[H]
  
  \includegraphics[width=0.85\textwidth]{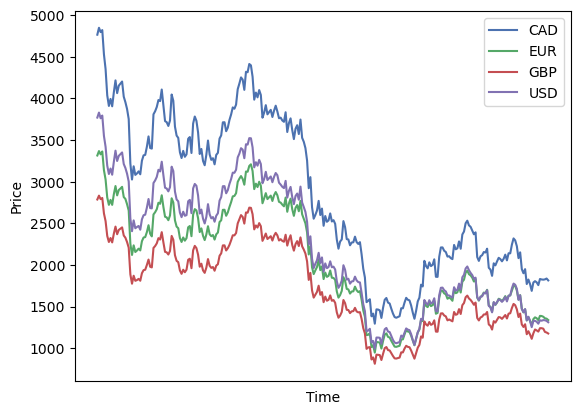}
  \caption{The price of ETH in multiple currencies.}
  \label{fig:volatility}
\end{figure}

Similar to gold, ETH has a global market against multiple fiat currencies. ETH can be traded independently with most major fiat currencies. And the unsynchronization of ETH against different fiat currencies is the inefficiencies in the market that provide trading opportunities. If a large volume of ETH is bought in USD, the value of ETH against USD will appreciate relative to ETH against GBP and CAD. This value discrepancy is the source of profit. There are some advantages of picking ETH as an anchor over gold because it is completely fungible and has low international transaction frictions. As shown in Figure~\ref{fig:volatility}, the CAD, EUR, GBP, and USD follow a similar trend but do not have 100\% correlation. Our technique aims to capture minor volatility movements and arbitrage market inefficiencies.

Though the cryptocurrency system operates in a decentralized manner, users can deposit their cryptocurrency into a centralized exchange for trading. The exchange functions similarly to stock or forex exchanges, allowing users to freely trade cryptocurrency for another cryptocurrency or fiat currency, either on spot or margin trading. Due to the decentralized nature of participants, cryptocurrency exchanges operate 24/7 with liquidity coming from global participation. These properties have led to experiments on cryptocurrency exchanges since as early as 2014~\citep{shah_bayesian_2014, brandvold_price_2015, cheah_speculative_2015}.

However, it is important to caution about the risks associated with cryptocurrency investment. \cite{kerr_cryptocurrency_2023} state that cryptocurrency carries a significant risk of financial fraud. Both \cite{kerr_cryptocurrency_2023} and \cite{pakhnenko_cryptocurrency_2023} compare cryptocurrencies to financial bubbles, highlighting the potential for significant market volatility and instability. This paper does not encourage investment in cryptocurrencies; rather, we only use them as intermediates for arbitrage in pair trading.

\section{Methodology}
\label{sec:opti-cont}

There are two stages in our trading technique. The first stage involves screening the assets. We select assets that are correlated and cointegrated with each other for a trading bucket. Once the bucket is formed, we monitor the differences between each asset and conduct trades based on opening and closing trading signals.

\subsection{Screening}
\label{sec:screening}

\begin{figure}[H]
  
  \includegraphics[width=0.85\textwidth]{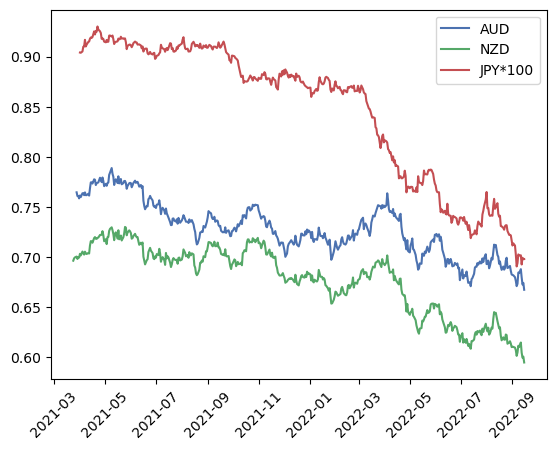}
  \caption{The volatility of forex rate against USD.}
  \label{fig:fx}
\end{figure}

In the screening stage, we analyze historical data to ensure the assets are suitable for arbitrage. Since movements are impacted by the volatility of both cryptocurrency and fiat currency, severe inflation or deflation in a fiat currency could mislead our trading signals. Therefore, one major criterion is that the volatility of fiat currencies should be small compared to that of cryptocurrency. An example of strong fiat depreciation is illustrated in Figure \ref{fig:fx}: the difference between AUD and NZD is relatively stable compared to the rapid drop of JPY, which would send a short signal based on its value change without mean reversion. To select appropriate fiat currencies for the trading bucket, we reviewed the historical trends of fiat currencies to ensure they have the desired statistical relationships.

The first desired statistical property is \textit{correlation}. We examine correlation using Pearson's correlation coefficient \citep{cohen_pearson_2009}, which ranges between $[-1, 1]$. The higher the absolute value of the coefficient, the stronger the association between the two variables. A coefficient of $-$1 represents an absolute negative association, 1 represents an absolute positive association, and 0 indicates no association between datasets.

The second desired statistical property is \textit{cointegration}. We refer to the price difference between two assets as \textit{spread}. In the long term, a widening spread exposes arbitrageurs to bankruptcy risk. Therefore, the spread between two assets must have a constant mean and variance. A time-series dataset with these characteristics is called \textit{stationary}. Cointegration describes the relationship between two time-series datasets whose difference is stationary. Cointegration is vital in pair trading because forming a pair with cointegration in the cryptocurrency market means that longing one asset and shorting another effectively holds the price difference. If the spread is stationary with a constant mean, convergence should always follow divergence. This property is called \textit{mean reversion}. In this research, the Engle--Granger test \citep{engle_co-integration_1987} is used to examine the \mbox{cointegration relationship.}

\subsection{Long and Short Trading Signals}
\label{sec:long-short}

After selecting the assets, we need to establish rules for trading the multivariate assets. Key questions include: How do we ensure market neutrality for multiple assets? How should we define and react to trading signals? How do we allocate our principal capital?

\subsubsection{Multivariate Pairs}

We define a consistent notation for \textit{pairs}:

\begin{itemize}
    \item ETH/USD or ETH/CAD denote the price of ETH in USD or CAD, respectively. 
    \item A pair between ETH/USD and ETH/CAD, is referred to as ETH/USD:ETH/CAD, abbreviated as USD:CAD.
\end{itemize}

Theoretically,
 we can form a maximum of $k(k-1)/2$ pairs from $k$ assets.

\subsubsection{Buy/Sell Signaling}
\label{sec:signaling}
Arbitrage relies on anomalies. We extract anomalies by comparing the normalized spread with the standard deviation. We use two parameters, \textit{open\_threshold} and\textit{ close\_threshold}, to tune the trading signals. For a two-asset trading scenario, the trading process between assets $i$ and $j$ is as follows:

\begin{itemize}
    \label{ite:lns}
    \item Calculate the difference of the logarithmized \textit{price} to obtain the \textit{spread} array.
    \item Normalize the spread array using the \textit{z-score}.
    \item Open a new position if there is no existing position and the \textit{z-score} $>$ \textit{open\_threshold} $\times$ \textit{standard\_deviation}.
    \item Close the position if there is an open position and the \textit{z-score} $<$ \textit{close\_threshold} $\times$ \textit{standard\_deviation}.
    \item Repeat until the end of the arbitrage period.
\end{itemize}

We provide a graphical example to explain the proposed approach. Table~\ref{tab:longnshort} gives the statistics and graphs for a three-asset scenario. Since each asset moves independently, each pair has a different standard deviation after normalization. The spreads are volatile, with thresholds represented by auxiliary lines. Every touch of the \textit{open\_threshold} indicates a position-opening signal. Conversely, if an open position exists and the spread touches the \textit{close\_threshold}, it signals a position closing.

\begin{table}[H]
  \caption{\label{tab:longnshort} Spread trading signaling.}
\newcolumntype{R}{>{\raggedleft\arraybackslash}X}
    \begin{tabularx}{\textwidth}{r RRR}
      \multicolumn{4}{c}{\includegraphics[width=0.8\textwidth]{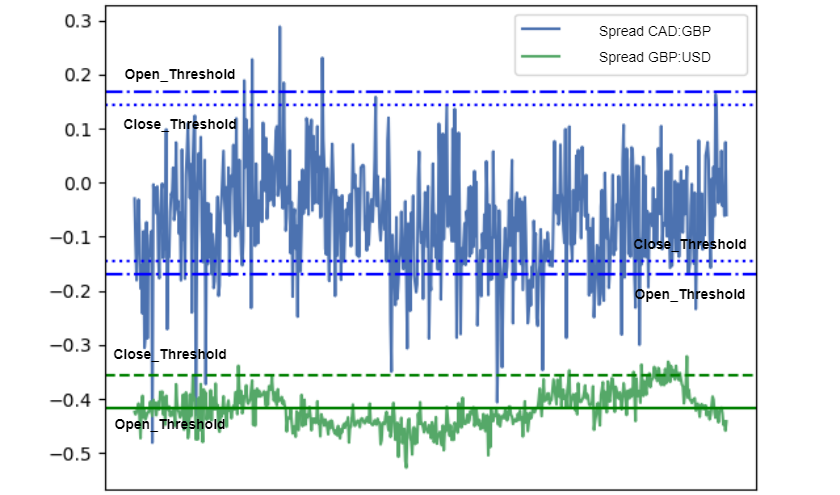}}\\[0.5em] 
      \toprule
      \textbf{Spread} & \textbf{Standard Deviation} & \textbf{Open\_Threshold} & \textbf{Close\_Threshold} \\
      \midrule
      USD:CAD & 0.024017 & 0.168122 & 0.144102 \\
      USD:GBP & 0.059545 & 0.416817 & 0.357270 \\
      CAD:GBP & 0.049811 & 0.348677 & 0.298866 \\
      \bottomrule
    \end{tabularx}
  \noindent{\footnotesize{In a cut-off period on 17 January 2021,
 a visualized spread-trading signaling based on the \textit{open\_threshold} as 7 and \textit{close\_threshold} as 6.}}
\end{table}
\subsubsection{Position Movements}

\begin{figure}[H]
    
    \includegraphics[width=0.8\textwidth]{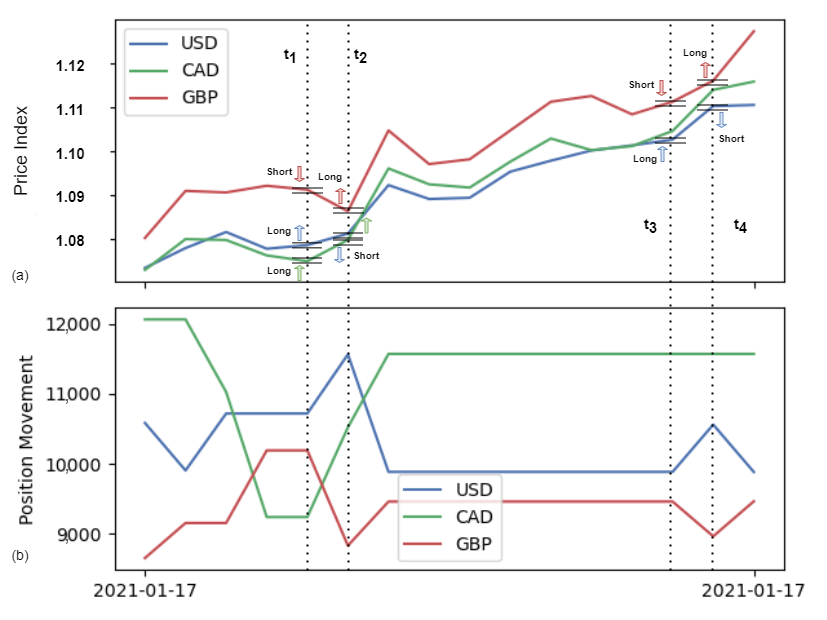}
    \caption{Actions
 and position movements for a three-currency scenario with plot {(a)} as price indices with trading positions and plot {(b)} as position movements over time.}
    \label{fig:longnshort}
\end{figure}

Suppose we return from the spreads to the actual asset granularity (Figure \ref{fig:longnshort}). The initial principal is distributed among the fiat currencies we are trading. Traditionally, we long the loser and short the winner for every positive signal. However, multiple spreads might produce conflicting signals. The example in Figure~\ref{fig:longnshort}a shows the normalized ETH price growth in USD, CAD, and GBP, while Figure~\ref{fig:longnshort}b tracks the principal movements of our fiat currency holdings.

\vspace{-3pt}

At time $t_1$, GBP starts a rapid decline. Both relevant spreads GBP:CAD and GBP:USD shrink, triggering trading signals. Our corresponding actions are longing ETH with GBP and shorting ETH to USD and CAD. At a later time $t_2$, there is a rebound the in ETH price against all currencies. Although all fiat currencies are rising, the USD reversal is weaker than the others, making the USD:CAD and USD:GBP spreads greater than the open threshold. This triggers the use of USD to buy ETH and sell ETH for CAD and GBP. By design, it is possible for a fiat currency to be longed and shorted simultaneously in different spreads. 

The classical scenario of two-asset trading still exists because each spread is monitored individually. For example, the movement at time $t_3$ does not trigger trading in CAD. The latency of rebound in USD makes only the \textit{z-score} of USD:GBP greater than the open threshold, so a position is opened for the spread USD:GBP. Later after time $t_4$, when the spread widens, the position opened at time $t_3$ is closed.

There are more benefits to including multivariate assets. If a fiat currency receives a long signal from one pair and an equivalent short signal from another, the actions can cancel out on paper, saving fees. Furthermore, when one fiat currency reveals a strong anomalous movement, all relevant pairs will display signals for the principal flowing in or out of that fiat currency. If there is an entry signal in USD:GBP, all the USD can be used to conduct the operation, ensuring no USD sits idle. Though ideally, this mitigation saves on transaction fees, signals that cancel each other are rare. More often, we have multiple short and long signals happening simultaneously. Furthermore, one short signal may come from a position closure and another from a new position opening. Therefore, we need to devise a mechanism to determine how funds should be distributed precisely.

\subsection{An Optimal Approach to Principal Allocation}
\label{sec:optimisation}

Resource distribution addresses how much to invest in each asset whenever a trading opportunity arises. At every trading interval, as long as a trading opportunity is detected, the optimization process is executed to make a quantitative decision on the amount to invest. Optimization is unnecessary if we only trade a two-asset pair since all the capital (or a pre-determined amount) can be allocated for the long and short positions. However, with more than two assets, the multilateral relationships necessitate distributing resources for each signal result to achieve the optimal outcome. The optimization process aims to maximize return with minimum risk under given constraints. In this research, a bi-objective optimization formulation inspired by the Markowitz portfolio allocation~\citep{markowitz_portfolio_1952} is employed to find the optimal investment solution.

\subsubsection{Parameters}

\begin{enumerate}

\item Let
 $\mathbf{c}$ be a vector of participating currencies, where $|\mathbf{c}| \in \mathbb{Z}^{+}$ represents the number of currencies.
  
\item Let $\mathbf{n}$ be a vector of pairs formed from currencies, where $|\mathbf{n}| \in \mathbb{Z}^{+}$ represents the number of pairs.

\item Let $\mathbf{t} \in [1, 2, 3, ..., T]$ represent the timestamp of execution, and $T$ denotes the total length of the arbitrage period.

\item Let $\mathbf{P} \in \mathbb{R}^{\mathbf{|c|}\times T}$ be a matrix representing ETH prices in $|\mathbf{c}|$ fiat currencies for the arbitrage period $T$.
$$
\begin{aligned}
  \begin{cases}
    \text{ETH/USD}:[P_{USD, t_1}, P_{USD, t_2}..., P_{USD, t_T}]\\
    \text{ETH/CAD}:[P_{CAD, t_1}, P_{CAD, t_2}..., P_{CAD, t_T}]\\
    \text{ETH/GBP}:[P_{GBP, t_1}, P_{GBP, t_2}..., P_{GBP, t_T}]\\
  \end{cases}
\end{aligned}
$$

\item Let $\mathbf{r} \in \mathbb{R}^{\mathbf{|c|}}$ records the average returns of the ETH price in $|\mathbf{c}|$ fiat currencies throughout the arbitrage period, and let expected profit $\mathbf{EP} \in \mathbb{R}^{|\mathbf{n}|\times 2}$ be a matrix for $\mathbf{|n|}$ pairs recording the average expected profit once an arbitrage position is opened. Since each spread consists of two different assets, the speeds of mean reversion should differ from pair to pair. Therefore, we measure a mean-reversion vector $\mathbf{mr} \in \mathbb{R}^{\mathbf{|n|}}$ to record the average reversion time to the mean from historical data. $\mathbf{EP}$ is calculated based on the combination of returns and mean-reversion speed as follows:
$$
\left[
  \begin{array}{c c} 
    mr_{USD,CAD}\times r_{USD} & mr_{USD,CAD}\times r_{CAD}\\ 
    mr_{USD,GBP}\times r_{USD} & mr_{USD,GBP}\times r_{GBP}\\
    mr_{CAD,GBP}\times r_{CAD} & mr_{CAD,GBP}\times r_{GBP}\\
  \end{array}
\right] 
$$ 

\item Let covariance $\mathbf{COV} \in \mathbb{R}^{\mathbf{|n|}\times2\times2}$ be a three-dimensional matrix. It contains $2\times2$ matrices for $\mathbf{|n|}$ pairs. The covariance matrix is leveraged to compute the risk of opening a position. An individual $2\times2$ matrix from the three-dimensional slice is a variance--covariance matrix for the two fiat currencies in a pair. An example of the $2\times2$ matrix is shown below:
$$
\begin{bmatrix}\text{var}({USD}) & \text{cov}(USD,CAD)\\\text{cov}(USD,CAD) & \text{var}(CAD)\end{bmatrix}
$$

\item Let transaction cost $tc \in \mathbb{R}^{\geq 0}$ be a constant, representing the percentage deducted from each investment amount as commission.

\item Let trading weights $\mathbf{tw} \in \mathbb{R}^{\mathbf{|c|}}$ be a vector representing the weight occupation of each currency. Since our arbitrage is a continuous process, every trade decision incorporates historical information. A strong buy signal guides the optimization process to invest all available USD into ETH. Without using margin trading, the maximum investment is 100\%. For example, if 40\% of USD is already used in an open position (long USD), the maximum available for the next signal is 60\%. This 40\% represents a weight that is actively being traded, hence the term trading weights, which can be positive or negative.

\end{enumerate}

\subsubsection{Variable}

Let Weights $\mathbf{W} \in \mathbb{R}^{N\times2}$ be the variable of the optimization in the form of a matrix for weights from every pair. The first column is the long currencies with positive weights, and the second column is the short currencies with negative weights. The negative weights capture the fact that the capital is reduced when going long, and the capital increases when going short. For instance:
$$
\left[
  \begin{array}{c c} 
    W_{USD} & W_{CAD}\\ 
    W_{USD} & W_{GBP}\\
    W_{CAD} & W_{GBP}\\
  \end{array}
\right] 
$$

\subsubsection{Objective}
\label{sec:objective}

The objective of the optimization is to maximize the return while minimizing the risk. We receive trading signals from spreads, so the decisions for investment weights are in pairs. When we need to long USD and short CAD in the pair USD:CAD, we are effectively applying the weight to bet on the rise of $r_{USD}$ and the drop of $r_{CAD}$. The expected profit is calculated as the weights multiplied by the expected profit, presented as:

\begin{equation}
\sum_{i=1}^{\mathbf{|n|}} W_{n} \cdot (\mathbf{EP}_{n} \odot [1, -1])^{'} \, , \, \forall n \in \mathbf{n}
\end{equation}
where $\odot$ is the Hadamard product, with long weights positive and short weights negative. We also aim to minimize risk, meaning the portfolio should have as little volatility as possible. According to Modern Portfolio Theory \citep{markowitz_portfolio_1952}, the portfolio volatility is $\boldsymbol{W} \times \boldsymbol{COV_{2\times2}} \times \boldsymbol{W}^{'}$ for any pair spread in $\mathbf{n}$. Since long weights are positive and short weights are negative, we need to take the Hadamard product with $\begin{bmatrix}1 & -1\\-1 & 1\end{bmatrix}$ to correctly capture the portfolio covariance in the covariance--variance matrix. Overall, for $\mathbf{n}$ pairs, we obtain Equation~(\ref{equ:risk}).

\begin{equation}
  \sum_{i=1}^{\mathbf{|n|}}W_{n} \cdot \mathbf{COV}_{n} \odot \begin{bmatrix}1 & -1\\-1 & 1\end{bmatrix} \cdot W_{n}^{'}\, , \, \forall n \in \mathbf{n}
  \label{equ:risk}
\end{equation}

\subsubsection{Constraints}
\label{sec:upper-bounds-lower}

The first constraint is that we can never invest more than what we have. Therefore, in every trading pair, the long weights cannot surpass 100\%, and the short weights shall not be less than $-$100\%:

\begin{equation}
  0 \leq W_{n,long} \leq 1, \; -1 \leq W_{n,short} \leq 0, \; \forall n \in \mathbf{n}
  \label{equ:bounds}
\end{equation}

Another constraint comes from the weight at the currency level. Every currency should not be spent more than 100\% at any time of the trade. Every time the optimization process runs, it calculates the summation of the current trading weight, long weight, and short weight to ensure that it does not exceed 100\% for each individual currency.

\begin{equation}
  tw_c + \sum_{n \in \mathbf{n}} W_{c,long} - \sum_{n \in \mathbf{n}} W_{c,short} \leq 1\, ,\, \forall c \in \mathbf{c}
  \label{equ:weights_tc}
\end{equation}

Market neutrality is the ability to generate stable income regardless of the market condition, and it is one of the major incentives for investors to select pair trading from other quantitative trading methods. Market neutrality requires an equivalent distribution of investment in the long and short positions. In that, we need to translate the weights back to the amount of ETH. The price of currency $c$ at time $t$ is attainable as $P_{c,t}$. In our case, whenever an ETH is bought in one currency, that ETH should
be sold instantaneously. Hence, we expect the last constraint is that every pair position reflecting back on assets is equalized in the unit of ETH. Transaction cost tweaks the balance slightly. The amount of ETH received from longing is reduced because of transaction costs. However, the amount of ETH required to obtain a certain amount of fiat currency increases. The market neutrality constraint is presented in Equation~(\ref{equ:neutrality_tc}):

\begin{align}
  \label{equ:neutrality_tc}
   \nonumber
    -W_{n,short}\cdot (1+tc)P_{c_1,t}= W_{n,long}\cdot (1-tc)P_{c_2,t}\\
    \forall t \in \mathbf{t},\; \forall (c_1, c_2)\in \mathbf{c} ,\;\forall n \in \mathbf{n}
\end{align}

\subsubsection{Optimization Equation}

Given the above, for the purpose of maximizing profitability and minimizing risk, one can optimize the allocation of the principal capital by solving the bi-objective optimization problem in Equation~(\ref{equ:portfolio_optimisation}). The ultimate goal of the optimization process is to calculate a set of quantitative weights when opening any position on the pair spreads.

We understand that market participants have different tendencies toward risk aversion. Therefore, we use a coefficient $\lambda$ (default to 1) to measure the degree of risk aversion in the objective. The greater the $\lambda$, the greater the risk aversion. In extreme cases, $\lambda = 0$ stands for risk neutrality. The $\lambda$ and portfolio variance form a penalty factor in the objective to cater to investors' risk preferences.

\begin{align}
  \begin{cases}
    \text{obj}: 
        & \sum_{i=1}^{\mathbf{|n|}} W_n \cdot (\mathbf{EP}_n \odot [1, -1])^{'} \\
        & \quad - \lambda \sum_{i=1}^{\mathbf{|n|}} W_n \cdot \mathbf{COV}_n \odot \begin{bmatrix}1 & -1\\-1 & 1\end{bmatrix} \cdot W_n^{'}, \, \forall n \in \mathbf{n} \\
    s.t. & 0 \leq W_{n,long} \leq 1, \; -1 \leq W_{n,short} \leq 0, \; \forall n \in \mathbf{n} \\\\
    s.t. & tw_c + \sum_{n \in \mathbf{n}} W_{c,long} - \sum_{n \in \mathbf{n}} W_{c,short} \leq 1,\, \forall c \in \mathbf{c} \\
    s.t. & -W_{n,short} \cdot (1+tc) P_{c_1,t} = W_{n,long} \cdot (1-tc) P_{c_2,t} \\
         & \quad \forall t \in \mathbf{t}, \, \forall (c_1, c_2) \in \mathbf{c}, \, \forall n \in \mathbf{n}
  \end{cases}
  \label{equ:portfolio_optimisation}
\end{align}
where:
\begin{itemize}
  \item $\mathbf{n}$: set of all pairs;
  \item $\mathbf{W}_n$: weight vector for the n-th pair;
  \item $\mathbf{EP}_n$: expected profit vector for the n-th pair;
  \item $\odot$: Hadamard product (element-wise multiplication);
  \item $\lambda$: risk aversion coefficient;
  \item $\mathbf{COV}_n$: covariance matrix for the n-th pair;
  \item $W_{n,long}$: long weight for the n-th pair;
  \item $W_{n,short}$: short weight for the n-th pair;
  \item $tw_c$: trading weight for currency $c$;
  \item $\mathbf{c}$: set of all currencies;
  \item $tc$: transaction cost;
  \item $P_{c,t}$: price of currency $c$ at time $t$;
  \item $\mathbf{t}$: set of all time periods.
\end{itemize}

\section{Simulation}
\label{sec:simulation}

This section describes the process and structure of our simulation (Algorithm \ref{alg:sim}) on historical data. The process consists of multiple stages: \textit{screening}, \textit{hist\_analysis}, \textit{trading}, and \textit{optimization}.

The \textit{screening} (cf. Section~\ref{sec:screening}) determines which assets should be included within our trading bucket. ETH pricing data are extracted from crypto exchanges against fiat currencies as the input (Algorithm~\ref{alg:sim} line~\ref{alg:pre}). We choose the trading period by splitting the pricing data into formation data and trading data. In the formation period, ETH prices against fiat currencies are screened to fill the bucket with fiat currencies that have strong correlations and high cointegration test passing rates. Once we drop the fiat currencies with weak correlation/cointegration relationships, the screened currencies are included in \textit{participating\_assets}. Then, we call the function \textit{hist\_analysis}.

In the \textit{hist\_analysis} process (Algorithm~\ref{alg:sim} Line~\ref{alg:tra}), standard deviation, expected return, and expected risk are computed using the historical formation data. Parameters \textit{open\_threshold} and \textit{close\_threshold} are computed using the formation data (cf. Section~\ref{sec:signaling}). Knowing when to trade is not sufficient; we also need to calculate how much to bid/ask on each order. The \textit{trading} function (Algorithm \ref{alg:sim} line~\ref{alg:qua}) simulates by amount. The first step is to set up $orig\_amount$, which is the initial amount of each participating fiat currency at the beginning of the trade. Secondly, every two currencies are combined into spreads. Next, we loop through all the trading times and constantly check if any open position meets the closing criteria:

\begin{itemize}
\item If yes, we close the open position to harvest profit. The trading weight vector $\mathbf{tw}$ is also updated as described in Section~\ref{sec:optimisation}.
\item If no, we leave it as is and check again in the next time sample.

\end{itemize}

\begin{algorithm}[H]

  \DontPrintSemicolon
  \caption{Arbitrage
 Technique Simulation}
  \label{alg:sim}
  \newcommand\mycommfont[1]{\footnotesize\ttfamily\textcolor{black}{#1}}
  \SetCommentSty{mycommfont}
  \SetNoFillComment
  \SetKwFunction{Opt}{optimisation}
  \SetKwFunction{Qua}{trading}
  \SetKwFunction{Tra}{hist\_analysis}
  \SetKwFunction{Pre}{screening}
  \SetKwProg{Fn}{def}{:}{}
  \KwData{Pricing\_Data} 
  \tcc{Historical pricing data for cryptocurrency against fiat currencies}
  \BlankLine
  \Fn{\Opt{trading\_data, expected\_return, expected\_risk \label{alg:opt}}}{
    invest\_weights $\leftarrow$ trading\_data, expected\_return, expected\_risk\;
    \tcc{Optimize investment weights from historical data using Equation~(\ref{equ:portfolio_optimisation})}
    \KwRet invest\_weights\;
  }
  \BlankLine
  \Fn{\Qua{open\_date, close\_date, trading\_data, open\_threshold, close\_threshold, expected\_return, expected\_risk} \label{alg:qua}}{
    orig\_amount()\;
    \tcc{Set the starting amount of each asset}
    spread\_data $\leftarrow$ trading\_data\;
    \tcc{Calculate spread data from asset trading period pricing data}
    \For{time \textbf{in} [open\_date, close\_date]}{
      \If{$\left|\text{spread\_data[time]}\right| < $ close\_threshold \& position\_open = True}{
        close\_position()\;
        \tcc{Close position for a pair when there is an open position and the spread is smaller than the closing threshold}
      }
      \If{$\left|\text{spread\_data[time]}\right| > $ open\_threshold \& position\_open = False}{
        invest\_weights $\leftarrow$ \Opt{trading\_data, expected\_return, expected\_risk}\;
        open\_position(invest\_weights, orig\_amount)\;
        \tcc{Open position when the spread has no current position and is greater than the opening threshold}
      }
    }
  }
  \BlankLine
  \Fn{\Tra{formation\_data, trading\_data \label{alg:tra}}}{
    std\_dev, expected\_return, expected\_risk $\leftarrow$ formation\_data\;
    \tcc{Calculate standard deviation, expected risk/return among spreads}
    open\_threshold, close\_threshold $\leftarrow$ std\_dev, formation\_data \;
    \tcc{Set trading thresholds between spreads}
    \Qua{open\_date, close\_date, trading\_data, open\_threshold, close\_threshold, expected\_return, expected\_risk}\;
    \tcc{Call trading function to trade from open\_date to close\_date}
  }
  \BlankLine
  \Fn{\Pre{pricing\_data\label{alg:pre}}}{
    formation\_data, trading\_data $\leftarrow$ pricing\_data \label{alg:preprocess_start}\; 
    \tcc{Split the data into formation period and trading period}
    assets\_correlation, assets\_cointegration $\leftarrow$ formation\_data\;
    \tcc{Calculate the correlation and cointegration between assets}
    participating\_assets $\leftarrow$ assets\_correlation, assets\_cointegration\;
    \tcc{Select assets above threshold}
    formation\_data, trading\_data $\leftarrow$ formation\_data, trading\_data, participating\_assets \label{alg:preprocess_end}\;
    \tcc{Filter the data to only participating assets}
    \Tra{formation\_data, trading\_data}\;
  }
  \Pre{Pricing\_Data}
\end{algorithm}
\vspace{12pt}

Similarly, all the spreads that do not have an ongoing open position need to be checked at every time sample to see if an open signal is triggered. Different from closing a position, open-position actions must consider how much $orig\_amount$ to spend on each open signal.

To calculate the investment amount, we call the \textit{optimization} function (Algorithm~\ref{alg:sim} line~\ref{alg:opt}). This function uses trading data and properties such as expected return and expected risk to determine the weights based on Equation~(\ref{equ:portfolio_optimisation}). Once the optimal weights are calculated, they are returned to the \textit{trading} function to continue position opening until the next call.

\section{Experiment}
\label{sec:results}

In this section, we apply our technique on historical ETH datasets against USD, EUR, CAD, and GBP. We select three market conditions to test the applicability of the \acrlong{ott} (cf. Algorithm~\ref{alg:sim}), including a bull-run market (1 January 2021 to 1 January 2022, 365 days), a bear-run market (1 January 2022 to 1 January 2023, 365 days), and a full-cycle market (15 January 2021 to 1 October 2022, 625 days) for trading intervals 1 min, 5 min and 60 min.

\subsection{Experimental Setup}
\label{sec:experimental-setup}

An overarching view of our method is to pair trade fiat currencies with a cryptocurrency as an intermediate. We adopt the trading bucket from the Kraken
\endnote{Kraken exchange (\url{https://www.kraken.com/}).} exchange because it provides freely accessible asset trading against multiple fiat currencies for individual researchers. Kraken is also one of the top exchanges offering the most fiat options for platform users without geographical restrictions. 

We trade a bucket of fiat currency pairs referenced to ETH for several reasons: 
\textcircled{1}~Cryptocurrency markets have substantial liquidity and depth;
\textcircled{2} the volatility in the cryptocurrency market overwhelms the inherent volatility in fiat currencies, providing opportunities for arbitrage.
\textcircled{3} for fiat currency pairs (equivalently spreads) referenced to cryptocurrencies, ETH has a strong cointegrated and stationary relationship (cf. Table~\ref{tab:corr});

Currencies with the most consistent trading volumes available on the Kraken exchange are USD, CAD, GBP, and EUR. The period from 15 January 2022 to 1 October 2022 represents a perfect full-cycle market where ETH prices against USD climb roughly quadruple in the first half and fall back to the beginning price in the second half. For more verifications, we also examined a whole-year bull market (1 January 2021 to 1 January 2022) and a whole-year bear market (1 January 2022 to 1 January 2023).

\begin{table}[H]
  \caption{\label{tab:corr} ETH correlation and cointegration among currency pairs.}
    \begin{tabularx}{\textwidth}{r CCCCCC}
      \toprule
      \multirow{2.5}{*}{\textbf{Interval}} & \multicolumn{2}{c}{\textbf{1 min}} & \multicolumn{2}{c}{\textbf{5 min}} & \multicolumn{2}{c}{\textbf{60 min}} \\
        \cmidrule{2-7}
      ~ & \textbf{Corr} & \textbf{Coint} & \textbf{Corr} & \textbf{Coint} & \textbf{Corr} & \textbf{Coint} \\
      \midrule
      USD:CAD & 0.978 & (77.8\%) & 0.998 & (100.0\%) & 0.965 & (89.9\%) \\ 
      USD:GBP & 0.916 & (66.7\%) & 0.974 & (94.1\%) & 0.908 & (72.5\%) \\ 
      USD:EUR & 0.997 & (77.8\%) & 0.914 & (72.5\%) & 0.994 & (69.6\%) \\ 
      CAD:GBP & 0.917 & (85.2\%) & 0.991 & (72.5\%) & 0.903 & (81.2\%) \\ 
      CAD:EUR & 0.978 & (88.9\%) & 0.980 & (94.1\%) & 0.969 & (98.6\%) \\ 
      GBP:EUR & 0.917 & (88.9\%) & 0.921 & (98.0\%) & 0.913 & (91.3\%) \\
      \bottomrule
    \end{tabularx}
   \noindent{\footnotesize{This table provides average rolling correlation calculations and cointegration test results for each currency pair between 1 January 2018 and 1 January 2020.}}
\end{table}

The datasets we use are publicly available from the Kraken OHLCVT dataset\endnote{Please refer to the Data Availability Statement for data access. 
}, which includes historical trading data from Kraken exchange for open price, highest price, lowest price, close price, trading volume, and number of trades at different sampling time intervals.

\subsubsection*{Accounting for Transaction Fees}
\label{sec:transaction-cost}

Transaction costs in quantitative trading are always a non-negligible factor. One of the incentives to introduce optimal trading is the benefit of reducing transaction costs. The buy-and-hold strategy is least affected by transaction costs.

\textls[-15]{Kraken Exchange uses a volume-based fee scheme, where higher trading volumes result in lower transaction fees. In practice, quantitative trading typically involves large trading volumes. For our analysis, we adopt a flat 0.1\% transaction fee for both long and short orders.}

\subsection{Data Source Screening}
\label{sec:data-source-scre}

As mentioned previously (cf. Section~\ref{sec:screening}), a preliminary requirement of the proposed \acrshort{ott} is that fiat currencies quoted for ETH are correlated and cointegrated. If not, it indicates that trading on the bucket of ETH/fiat pairs is unlikely to follow a mean-reversion process. We use the SPOT ETH price quoted in USD, CAD, GBP, and EUR from 1 January 2018 to 1 January 2020 for screening and formation (Algorithm~\ref{alg:sim} line~\ref{alg:pre}). We test the stationarity of the pairs (spreads) not only for the complete formation period but also for shorter cycles corresponding to the time frames of our arbitrage. Therefore, we examine the data on a rolling 1-week basis to test for correlation and stationarity via cointegration.

The correlation and cointegration test results are presented in Table~\ref{tab:corr}. USD:CAD under 1-week represents Pearson's correlation for USD:CAD between 1 January 2018 and 1 January 2020, divided into 105 weeks. The correlation presented is the simple weekly average for these 105 weeks. We expect our trading targets to be stationary, so we also expect the spreads we trade to be cointegrated during the period we hold the positions. Similar to Pearson's correlation test, the data are also divided into weeks for the Engle--Granger test. The passing percentages are presented in the table. We can see that the data generally have good correlations and cointegration relationships, allowing us to confidently conduct the proposed arbitrage.

\subsection{Tuning Trading Parameters}
\label{sec:tuning}

Once the stationarity of the spreads is examined, we compute several parameters, including thresholds, which determine the boundaries to open and close positions based on the formation data (Algorithm~\ref{alg:sim} line~\ref{alg:tra}).

\begin{figure}[H]
  \includegraphics[width=0.58\textwidth]{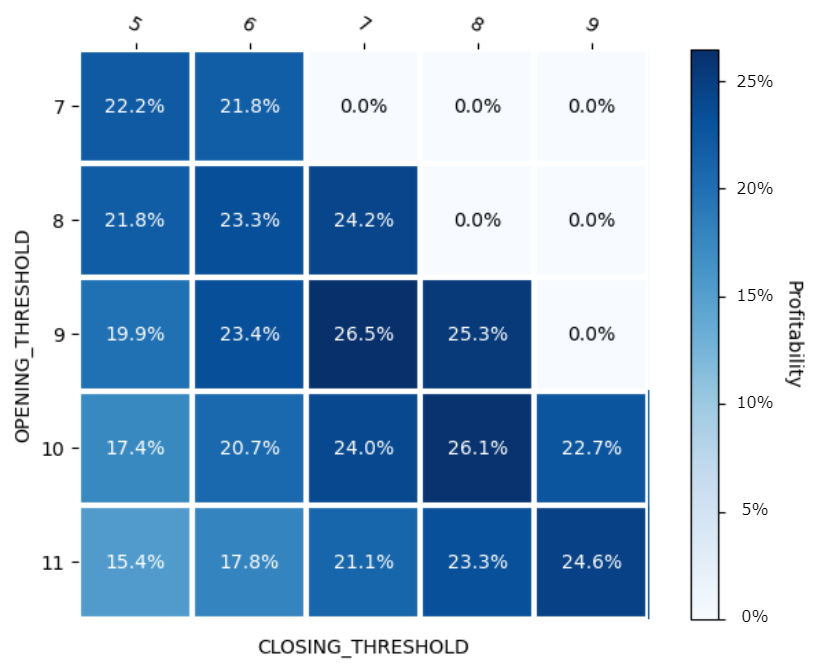}  
  \caption{Heatmap for profitability tuning under 5min interval between 15 January 2020 and 1 October 2022.}
  \label{fig:tuning}
\end{figure}

\textls[-15]{The two parameters that determine the closing and opening boundaries are \textit{open\_threshold}} and \textit{close\_threshold} (cf. Table~\ref{tab:longnshort}). In our research, we conduct a grid search for the thresholds that are likely to generate high profitability. Figure~\ref{fig:tuning} illustrates the grid search results for the 5min sample interval formation dataset. It is a heatmap for the indicators of profitability (calculation method in Section~\ref{sec:prof_cal}). A darker shade means better profit for the \textit{open\_threshold}/\textit{close\_threshold} combination. For the 5 min sampled dataset, opening at nine~standard deviations and closing at seven standard deviations generates the best profitability. The grid search is extended to the trading intervals of 1 min and 60 min as well. Table~\ref{tab:tuning} records the most lucrative combination for the trading intervals adopted in this research.

\vspace{-6pt}
\vspace{-6pt}
\begin{table}[H]
  \caption{\label{tab:tuning} The most profitable trading thresholds.}
\newcolumntype{R}{>{\raggedleft\arraybackslash}X}
    \begin{tabularx}{\textwidth}{r RR}
      \toprule
      \textbf{Interval} & \textbf{OPEN\_THRESHOLD} & \textbf{CLOSE\_THRESHOLD}\\
      \midrule
      1 min & 11 Standard Deviations & 9 Standard Deviations \\ 
      5 min & 9 Standard Deviations & 7 Standard Deviations\\ 
      60 min & 7 Standard Deviations & 6 Standard Deviations \\ 
      \bottomrule
    \end{tabularx}
   \noindent{\footnotesize{ The thresholds are calculated from retrospective grid searches from 05 January 2020 to 1 October 2022 for the most lucrative combination of \textit{open\_threshold} and \textit{close\_threshold}.}}
\end{table}

Another tunable parameter is the risk-preference parameter $\lambda$, which is used to adjust the penalty factor for investment risk in Equation~(\ref{equ:risk}). We can tweak $\lambda$ to cater to different risk preferences. A high $\lambda$ represents a strong penalty factor for portfolio volatility, leading to a more conservative investment decision. By default, $\lambda$ is set to 1, but we will also examine the profitability of arbitrage under a riskier $\lambda=0.5$ and a more conservative $\lambda=2$.

\subsubsection*{Calculating P\&L in Base Currency}
\label{sec:prof_cal}

Since we are trading with multiple fiat currencies, we need a unified method to measure profitability (or loss)—so-called P\&L. We choose a base currency among the participating assets, which is USD in our case. We obtain the foreign currency exchange rates for the beginning of the arbitrage and the end. Note that our testing dataset does \textit{not} overlap with the formation period (12 March 2016 to 31 December 2021). Hence, all our arbitrage tests are on out-of-sample (never seen before) datasets.

We start with USD, CAD, GBP, and EUR 10,000 each, totaling USD 42,340.72 on 15 January 2021. By 1 October 2022, if all the assets remain the same, their total value would be equivalent to USD 40,564.62 (Table~\ref{tab:prof_cal}). This calculation indicates a 4.19\% loss due to currency exchange rate changes. However, this does not necessarily mean a 4.19\% overall loss in action. Our technique tracks forex volatility as part of the trading signal. When USD is bullish, the strategy directs us to long ETH with USD against other currencies. Therefore, our profitability results from a compounded effect of cryptocurrency arbitrage and forex volatility arbitrage.

\begin{table}[H]
    \caption{\label{tab:prof_cal} Profit calculation from 15 January 2021 to 1 October 2022.}
    \newcolumntype{R}{>{\raggedleft\arraybackslash}X}
    \begin{tabularx}{\textwidth}{r RRRR}
    \toprule
        ~ & \multicolumn{2}{c}{\textbf{15 January 2021}} & \multicolumn{2}{c}{\textbf{1 October 2022}} \\ \midrule
        USD & USD 10,000 & (USD 10,000.00) & USD 12,000 & (USD 12,000.00) \\
        CAD & CAD 10,000 & (USD 7896.06) & CAD 8000 & (USD 5783.40) \\
        GBP & GBP 10,000 & (USD 13,186.23) & GBP 9000 & (USD 10,035.00) \\
        EUR & EUR 10,000 & (USD 11,258.42) & EUR 13,000 & (USD 12,746.22) \\
        \midrule
        Total & & USD 42,340.72 & & USD 40,564.62 \\
    \bottomrule
    \end{tabularx}
\end{table}

\subsection{Profit and Loss Results}
\label{sec:prof}

The proposed \acrfull{ott} experiments with out-of-sample datasets under different conditions to examine the validity of the technique, including full-cycle market, bull-run market, and bear-run market in this section. Further experiments with applications on BTC and SOL are included in Appendix \ref{sec:sensitivity}.

\subsubsection{Full-Cycle Market}
\begin{figure}[H]
  
  \includegraphics[width=0.7\textwidth]{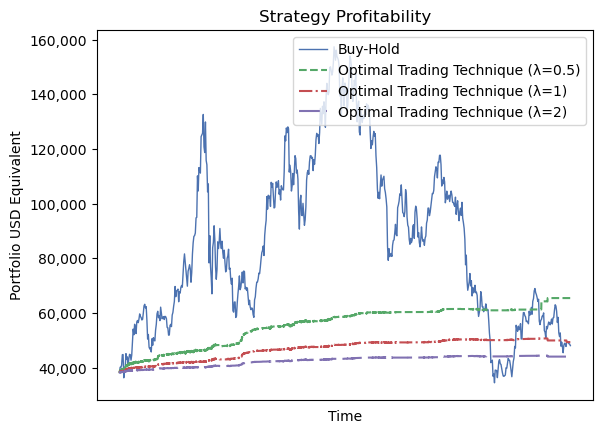}
  \caption{The
 profitability of full-cycle market between 15 January 2021 and 1 October 2022.}
  \label{fig:fullcycle}
\end{figure}

We use the method in Section~\ref{sec:prof_cal} to calculate the profitability in USD, presented in Table~\ref{tab:results}. The numbers in the table for various currencies represent the profit (or loss) at the end of the trading period, starting with 10,000 units of each currency. All currencies are converted to USD based on the exchange rate at the start of the trading period on 15 January 2021. We convert the currencies back to USD for profitability calculation on 1 October 2022. During this period, cryptocurrency experienced both a bull run and a bear run, allowing us to test our technique through peak and trough. The results of our technique are encouraging (Figure~\ref{fig:fullcycle}). We achieved an annualized return of 14.87\% in USD for the 1min sampled dataset ($\lambda$ = 1). For the 5min and 60min sampled datasets, the annualized returns were 15.49\% and 5.00\%, respectively, which greatly exceed the Distance Method (DM) in all experimented cases.

\begin{table}[H]
    \caption{\label{tab:results} Annualized profitability results in full-cycle market.}
\newcolumntype{R}{>{\raggedleft\arraybackslash}X}
      \begin{tabularx}{\textwidth}{l RRRRR}
        \toprule
        \textbf{Interval} & \textbf{BH} & \textbf{DM} & \textbf{\boldmath{OTT$_{\lambda=0.5}$}} & \textbf{\boldmath{OTT$_{\lambda=1}$}} & \textbf{\boldmath{OTT$_{\lambda=2}$}} \\ 
        \midrule
        1 min (TC 0.1\%) & 12.91\% & 5.33\% & 35.54\% & 14.87\% & 7.44\%  \\
        1 min (TC 0.0\%) & 13.14\% & 0.73\% & 79.98\% & 40.03\% & 19.97\% \\ 
        5 min (TC 0.1\%) & 12.91\% & 0.60\% & 37.74\% & 15.49\% & 7.74\% \\
        5 min (TC 0.0\%) & 13.14\% & 4.08\% & 90.03\% & 44.98\% & 22.49\% \\ 
        60 min (TC 0.1\%) & 12.91\% & 3.93\% & 9.93\% & 5.00\% & 2.51\% \\ 
        60 min (TC 0.0\%) & 13.14\% & 5.54\% & 37.43\% & 18.96\% & 9.49\%  \\
        \bottomrule
    \end{tabularx}
\noindent{\footnotesize{The
 arbitrage period is 15 January 2021 to 1 October 2022. Participating assets are USD, CAD, GBP, and EUR against ETH. The abbreviations used in the table are BH (Buy--Hold), DM (Distance Method), TC (Transaction Cost), and OTT (Optimal Trading Technique). \textit{open\_threshold} and \textit{close\_threshold} are tuned to the most lucrative combination.
}}
\end{table}

\vspace{-12pt}

We also conduct experiments on conditions where investors have a higher or lower tolerance for portfolio volatility. We observe a clear relationship: the lower the $\lambda$, the higher the profitability. This relationship meets our expectation that investors receive a risk premium in exchange for a more volatile portfolio.

Another comparison includes transaction costs. We can see that even a 0.1\% transaction cost significantly impacts our returns. On the surface, transaction cost charges a flat percentage of trading capital and influences the optimization process's expectation. Our technique trades on a frequent basis; hence, we can see the profitability is almost halved in three examined $\lambda$ values after including transaction costs. The Distance Method, on the other hand, does not have a significant impact due to transaction cost, because it trades less frequently (Table in Section~\ref{sec:indicators}
).

\subsubsection{Bull Market}

Financial assets are subject to fluctuations, with periods of both growth and decline. However, the duration of a trading period may be one-sided. The year 2021 was particularly noteworthy for cryptocurrencies, with Ethereum (ETH) appreciating fourfold across all major fiat currencies. Although such growth is unlikely to recur frequently, it is possible to evaluate trading techniques in extreme market conditions to determine if any specific trading traits emerge.

\begin{figure}[H]
  
  \includegraphics[width=0.7\textwidth]{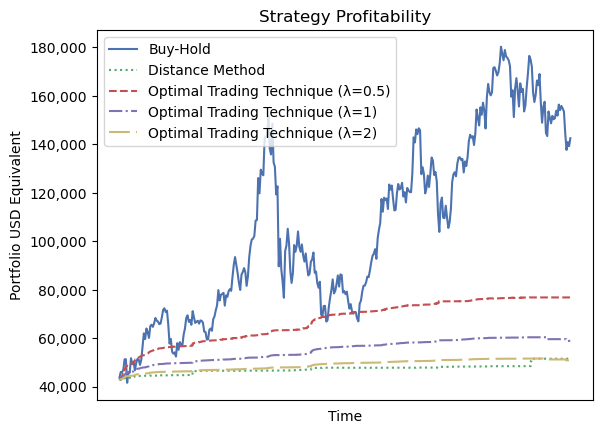}
  \caption{The
 profitability of bull market between 1 January 2021 and 1 January 2022.}
  \label{fig:bull}
\end{figure}

Table~\ref{tab:results_bull} presents the trading results for 2021. A significant growth trend is evident from the Buy--Hold strategy, which yielded a return of 412.09\%. Such a circumstance is rare and occurs only in exceptional cases. Compared to full-cycle market trading, both the Distance Method (DM) and our \acrfull{ott} yielded more profit in the bull market (Figure~\ref{fig:bull}). For a standard scenario of $\lambda=1$ and $TC=0.1\%$ over a five-minute period, the \acrshort{ott} achieved an annual return of 36.75\% (compared to 19.89\% for the DM under the same conditions). Single-sided markets are easier to estimate, and the expected\_return and expected\_risk are therefore closer to reality. The profitability trend is consistent with that of the full-cycle market. Arbitrages that do not entail transaction fees generate better returns than those with transaction fees. A lower value of $\lambda$ implies a higher acceptance of risk, resulting in a better overall arbitrage return.

\begin{table}[H]
    \caption{\label{tab:results_bull} Annualized profitability results in bull market.}
    \newcolumntype{R}{>{\raggedleft\arraybackslash}X}
    \begin{tabularx}{\textwidth}{l RRRRR}
        \toprule
        \textbf{Interval} & \textbf{BH} & \textbf{DM} & \textbf{\boldmath{OTT$_{\lambda=0.5}$}} & \textbf{\boldmath{OTT$_{\lambda=1}$}} & \textbf{\boldmath{OTT$_{\lambda=2}$}} \\ 
        \midrule
        1 min (TC 0.1\%) & 412.08\% & 22.17\% & 82.13\% & 39.85\% & 19.93\%  \\
        1 min (TC 0.0\%) & 412.90\% & 20.74\% & 170.04\% & 85.02\% & 42.51\% \\ 
        5 min (TC 0.1\%) & 412.08\% & 19.89\% & 77.78\% & 36.75\% & 18.38\% \\
        5 min (TC 0.0\%) & 412.90\% & 21.19\% & 165.34\% & 82.67\% & 41.34\% \\ 
        60 min (TC 0.1\%) & 412.08\% & 6.50\% & 25.23\% & 14.18\% & 6.40\% \\
        60 min (TC 0.0\%) & 412.90\% & 7.46\% & 45.25\% & 22.75\% & 11.37\%  \\
        \bottomrule
    \end{tabularx}
\noindent{\footnotesize{The
arbitrage period is 1 January 2021 to 1 January 2022. Participating assets are USD, CAD, GBP, and EUR against ETH. The abbreviations used in the table are BH (Buy--Hold), DM (Distance Method), TC (Transaction Cost) and OTT (Optimal Trading Technique). \textit{open\_threshold} and \textit{close\_threshold} are tuned to the most lucrative combination.
}}
\end{table}

\subsubsection{Bear Market}

\begin{figure}[H]
  
  \includegraphics[width=0.7\textwidth]{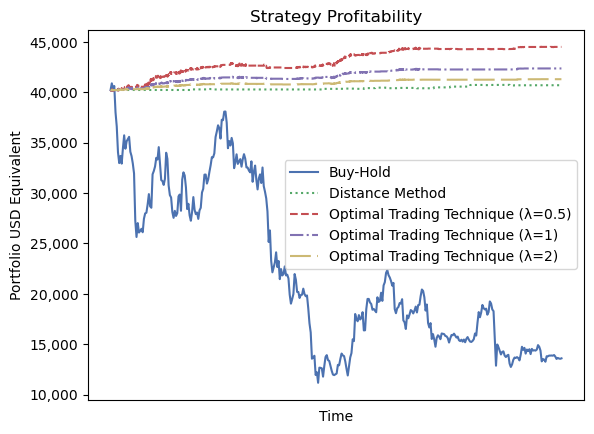}
  \caption{The
 profitability of bear market between 1 January 2022 and 1 January 2023.}
  \label{fig:bear}
\end{figure}

We also have an interest in examining how well our technique performs in a bear market (Figure~\ref{fig:bear}). Our optimization process calculates weights to avoid holding ETH in our portfolio. Therefore, in theory, the process should perform as well as in a bull market. Therefore, we experiment with the technique under the bear market in 2022, from 1 January 2022 to 31 December 2022. During this period, the price of ETH dropped by $-$61.77\%. We examine the profitability with data for this price dip in Table \ref{tab:results_downturn}.

\begin{table}[H]

    \caption{\label{tab:results_downturn} Annualized profitability results in bear market}
\newcolumntype{R}{>{\raggedleft\arraybackslash}X}
    \begin{tabularx}{\textwidth}{l RRRRR}
        \toprule
        \textbf{Interval} & \textbf{BH} & \textbf{DM} & \textbf{\boldmath{OTT$_{\lambda=0.5}$}} & \textbf{\boldmath{OTT$_{\lambda=1}$}} & \textbf{\boldmath{OTT$_{\lambda=2}$}} \\ 
        \midrule
        1 min (TC 0.1\%) & $-$61.84\% & 1.54\% & 6.58\% & 9.57\% & 1.64\%  \\
        1 min (TC 0.0\%) & $-$61.77\% & 1.71\% & 19.14\% & 3.29\% & 4.78\% \\ 
        5 min (TC 0.1\%) & $-$61.84\% & 1.86\% & 9.98\% & 4.99\% & 2.49\% \\
        5 min (TC 0.0\%) & $-$61.77\% & 2.08\% & 36.41\% & 18.20\% & 9.10\% \\ 
        60 min (TC 0.1\%) & $-$61.84\% & 5.25\% & 18.93\% & 9.01\% & 4.92\% \\
        60 min (TC 0.0\%) & $-$61.77\% & 5.53\% & 33.08\% & 13.93\% & 8.52\%  \\
        \bottomrule
    \end{tabularx}
\noindent{\footnotesize{The
arbitrage period is 1 January 2022 to 1 January 2023. Participating assets are USD, CAD, GBP, and EUR against ETH. The abbreviations used in the table are BH (Buy--Hold), DM (Distance Method), TC (Transaction Cost) and OTT (Optimal Trading Technique). \textit{open\_threshold} and \textit{close\_threshold} are tuned to the most lucrative combination.
}}
\end{table}

\vspace{-12pt}

During the bear market in 2022, ETH witnessed a significant drop in all major currencies. The traditional Buy--Hold strategy and most non-market-neutral techniques would experience a loss from the bear market. By not holding ETH at all, both the Distance Method and our \acrfull{ott} achieved positive returns in all experimented cases. We can see from Table~\ref{tab:results_downturn} that the patterns we found in a full-cycle market also \mbox{apply here.}

\subsubsection{Risk and Sharpe Ratio}
\label{sec:risk-sharpe-ratio}

Profit and risk are two sides of the same coin. We have shown that the lower the $\lambda$ parameter, the higher the profit. Risk is measured with the Sharpe Ratio~\cite{sharpe_mutual_1966}. The Sharpe Ratio $S_p$ is calculated as follows: 
$S_p = \frac{E[R_p] - R_f}{\sigma}$, where $R_p$ is the portfolio's annualized return, $R_f$ is the annual risk-free rate\endnote{We use 4\% based on the U.S. 10-Year Treasury Note at the time of writing.}, and $\sigma$ is the annualized standard deviation of the portfolio returns.

The results in Table~\ref{tab:risknsharpe} provide comparisons of different techniques. From the annualized standard deviation $\sigma$, we can see that \acrfull{dm} and \acrfull{ott} under all conditions have significantly lower volatility than Buy--Hold. DM is less volatile than OTT because it trades less. For OTT, there is an expected negative relation between $\lambda$ and $\sigma$. A greater value of $\lambda$ effectively reduces portfolio volatility. When measuring the overall portfolio performance with the Sharpe Ratio, lower $\lambda$ brings higher profitability and, therefore, a higher Sharpe Ratio.

\begin{table}[H]

    \caption{\label{tab:risknsharpe} Position movement for different trading strategies in markets.}
{\newcolumntype{R}{>{\raggedleft\arraybackslash}X}
      \begin{tabularx}{\textwidth}{l RRRRRR}
        \toprule
        ~ & \multicolumn{2}{c}{\textbf{Full-Cycle}} & \multicolumn{2}{c}{\textbf{Bull-Run}} & \multicolumn{2}{c}{\textbf{Bear-Run}} \\
        \cmidrule{2-7}
        \textbf{(Bull)} & \boldmath{$S_p$} & \boldmath{$(\sigma)$} & \boldmath{$S_p$} & \boldmath{$(\sigma)$} & \boldmath{$S_p$} & \boldmath{$(\sigma)$} \\
        \midrule
        Buy--Hold & 0.12 & (0.961) & 1.69 & (1.313) & $-$0.61 & (0.870) \\
        DM & $-$0.41 & (0.082) & 0.60 & (0.273) & $-$1.51 & (0.019) \\
        OTT$_{\lambda=2}$ & 0.44 & (0.104) & 2.11 & (0.070) & $-$0.32 & (0.042) \\
        OTT$_{\lambda=1}$ & 0.68 & (0.193) & 2.66 & (0.126) & 0.16 & (0.082) \\
        OTT$_{\lambda=0.5}$ & 1.11 & (0.343) & 4.34 & (0.174) & 0.41 & (0.161) \\
        \bottomrule
      \end{tabularx}}
\noindent{\footnotesize{The abbreviations used in the table are BH
(Buy--Hold), DM (Distance Method) and OTT (Optimal Trading Technique). $S_p$ is the Sharpe Ratio of the investment portfolio calculated based on a risk-free annual rate of 4\%. $\sigma$ is the standard deviation of the investment portfolio. And $\lambda$ represents the risk control factor. The full-cycle market is from 15 January 2021 to 1 October 2022. The bull market is from 1 January 2021 to 1 January 2022. The bear market is from 1 January 2022 to 1 January 2023.
}}
\end{table}
\subsection{Strategy Indicators}
\label{sec:indicators}
To have an overview of the trading strategy, we also look into the trading metrics, not just the profitability and risk. For consistency of the counting metrics, we set the following: one arbitrage equals two trades (position open/close) equals four orders (buy/sell).

Table~\ref{tab:indicators} includes some common trading strategy indicators. It consists of three sets of experiments: full-cycle, bull, and bear markets based on Section \ref{sec:prof}.

\begin{itemize}
\item \textit{number of trades}:
the total number of trades happened during the whole arbitrage period;
\item \textit{\% winning}: the percentage of winning trades among all trades;
\item \textit{\% losing}: the percentage of losing trades among all trades;
\item \textit{\% long wins}: the percentage of winning trades among all long trades;
\item \textit{\% short wins}: the percentage of winning trades among all short trades;
\item \textit{win/loss ratio}: the ratio of winning trades against losing trades;
\item \textit{average loss}: the average USD loss among losing trades;
\item \textit{average win}: the average USD win among winning trades;
\item \textit{largest loss}: the largest USD loss among losing trades;
\item \textit{largest win}: the largest USD win among winning trades;
\item \textit{avg holding hrs}: on average how many hours an arbitrage is open.
\end{itemize}

We can clearly see that our technique conducts trades much more frequently than the Distance Method. This is mainly because our spread focuses on the deviation of forex movements anchored to ETH, while the Distance Method concentrates on the distance. Our six experimental setups have a win/loss ratio greater than 1 (Table~\ref{tab:indicators}), forming the foundation of pair trading profitability. It is reasonable that the longer the interval, the fewer trades occur, resulting in longer average holding hours. We are also pleased to see that there are no particularly large values in the largest win or largest loss that could impact the overall revenue. This indicates the stability of our arbitrage strategy.

As for the comparison between the bull market and bear market, there is a significant decrease in the number of trades and a slight drop in the average profit (or loss) in the bear market. This indicates that although our strategy remains effective in a bear market, the quality and quantity of arbitrage opportunities are impacted.

\begin{table}[H] 

    \caption{\label{tab:indicators} Strategy Indicators.}
\newcolumntype{R}{>{\raggedleft\arraybackslash}X}
        \begin{tabularx}{\textwidth}{l RRRRRR}
        \toprule
        ~ & \multicolumn{3}{c}{\textbf{Optimal Trading Technique}} & \multicolumn{3}{c}{\textbf{Distance Method}}\\
\midrule
        Indicators (\textit{Full Cycle}) & 1 min & 5 min & 60 min & 1 min & 5 min & 60 min\\
        \midrule
        number of trades & 13,790 & 12,166 & 1906 & 112 & 56 & 30 \\ 
        \% of winning & 53.2\% & 54.4\% & 52.6\% & 63.3\% & 53.6\% & 56.7\% \\
        \% of losing & 46.8\% & 45.6\% & 47.3\% & 36.7\% & 46.4\% & 43.3\% \\ 
        \% of long win & 54.8\% & 55.5\% & 52.4\% & 78.6\% & 71.4\% & 73.3\% \\ 
        \% of short win & 51.6\% & 53.2\% & 53.0\% & 48.2\% & 35.7\% & 40.0\% \\ 
        win/loss ratio & 1.14 & 1.19 & 1.13 & 1.73 & 1.15 & 1.31\\ 
        average loss (US\$) & $-$\$6.09 & $-$\$8.67 & $-$\$37.31 & $-$\$50.72 &$-$\$174.64& $-$\$657.00\\ 
        average win (US\$) & \$7.00 & \$9.21 & \$37.43 & \$59.04 &\$210.41&\$518.33 \\ 
        largest loss (US\$) & $-$\$347.31 & $-$\$475.24 & $-$\$1078.51 & $-$\$299.56 &$-$\$864.43& $-$\$2806.60\\ 
        largest win (US\$) & \$386.01 & \$540.01 & \$976.62 & \$366.06 &\$862.34& \$2690.23\\ 
        avg holding hrs & 4.52 & 5.09 & 16.10 & 236.88 &565.50&933.07\\
        \midrule
        Indicators (\textit{Bull Run}) & 1 min & 5 min & 60 min & 1 min & 5 min & 60 min\\
        \midrule
        number of trades & 13,490 & 10,456 & 1620 & 400 & 290 & 122 \\ 
        \% of winning & 54.0\% & 55.2\% & 53.6\% & 62.5\% & 60.0\% & 58.2\% \\
        \% of losing & 46.0\% & 44.8\% & 46.4\% & 37.5\% & 40.0\% & 41.8\% \\ 
        \% of long win & 56.8\% & 57.1\% & 54.9\% & 74.0\% & 75.9\% & 70.5\% \\ 
        \% of short win & 51.3\% & 53.4\% & 52.3\% & 51.0\% & 44.1\% & 45.9\% \\ 
        win/loss ratio & 1.18 & 1.23 & 1.16 & 1.67 & 1.5 & 1.39\\ 
        average loss (US\$) & $-$\$6.38 & $-$\$9.61 & $-$\$48.33 & $-$\$53.67 &$-$\$78.53& $-$\$179.77\\ 
        average win (US\$) & \$7.97 & \$10.84 & \$48.89 & \$55.93 &\$84.80 &\$169.11 \\ 
        largest loss (US\$) & $-$\$335.44 & $-$\$804.18 & $-$\$1502.54 & $-$\$605.46 &$-$\$792.19 & $-$\$1193.86\\ 
        largest win (US\$) & \$359.88 & \$794.02 & \$1554.55 & \$635.35 &\$1480.52 & \$1325.57\\ 
        avg holding hrs & 2.88 & 4.03 & 19.83 & 41.99 & 56.79 & 114.67\\
        \midrule
        Indicators (\textit{Bear Run}) & 1 min & 5 min & 60 min & 1 min & 5 min & 60 min\\
        \midrule
        number of trades & 2146 & 3306 & 814 & 48 & 40 & 32 \\ 
        \% of winning & 52.6\% & 52.8\% & 54.3\% & 60.4\% & 55.0\% & 53.1\% \\
        \% of losing & 47.4\% & 47.2\% & 45.7\% & 39.6\% & 45.0\% & 46.9\% \\ 
        \% of long win & 47.2\% & 52.8\% & 60.2\% & 50.0\% & 50.0\% & 43.8\% \\ 
        \% of short win & 58.0\% & 52.9\% & 48.4\% & 70.1\% & 60.0\% & 62.5\% \\ 
        win/loss ratio & 1.11 & 1.12 & 1.19 & 1.53 & 1.22 & 1.13\\ 
        average loss (US\$) & $-$\$10.11 & $-$\$10.40 & $-$\$40.85 & $-$\$85.21 &$-$\$107.74& $-$\$153.77\\ 
        average win (US\$) & \$10.25 & \$10.51 & \$39.00 & \$73.48 &\$109.60&\$172.19 \\ 
        largest loss (US\$) & $-$\$420.11 & $-$\$737.76 & $-$\$1342.21 & $-$\$249.60 &$-$\$382.87& $-$\$350.41\\ 
        largest win (US\$) & \$440.61 & \$736.69 & \$1387.96 & \$294.02 &\$435.66& \$402.41\\ 
        avg holding hrs & 22.69 & 15.15 & 49.80 & 433.01 & 433.23 & 447.44\\
        \bottomrule
    \end{tabularx}
        \noindent{\footnotesize{The full-cycle market is from 15 January 2021 to 1 October 2022. The bull market is from 1 January 2021 to 1 January 2022. The bear market is from 1 January 2022 to 1 January 2023. All the data are collected under $\lambda$ as 1 and transaction cost as 0.1\%}}
\end{table}

\section{Conclusions}
\label{sec:conclusions}

In this paper, we examined the feasibility and profitability of multivariate pair trading. We used fiat currencies quoted for ETH as the bucket of our pairs. We constructed the spread focusing on anomalies in forex movements due to independent trades against cryptocurrencies. The trading threshold is a tunable parameter based on the standard deviation of spreads. We demonstrated a grid search result of profitability under untuned conditions. In addition to the conventional pair trading strategy, we also introduced an optimization process for resource allocation in multivariate pair trading, ensuring market neutrality and utilizing most of the holding assets without leverage.

The results are encouraging. The profitability of the proposed \acrfull{ott} is on average 2.72 times\endnote{The average calculation in the conclusion section is based on 1 min, 5 min, and 60 min sampled datasets for bull, bear, and full-cycle markets under a 0.1\% transaction fee.} more profitable compared to the baseline approach—the Distance Method. This implies that trading spread is more effective than distance in the volatile cryptocurrency market. Under the assumption of a 4\% risk-free interest rate, the Sharpe Ratio of the OTT achieved is a strong 1.17. For all the tested cases, the OTT achieved a Win/Loss ratio greater than 1, which drives steady capital growth.

Our bi-criteria optimization objective can be tuned with the risk tolerance parameter $\lambda$, which impacts the optimization process for how much risk a trader tolerates. For the full-cycle market, 5min sampled dataset, the high ($\lambda = 2$), intermediate ($\lambda=1$), and low ($\lambda=0.5$) risk aversion parameter values have annualized returns of 7.74\%, 15.49\%, and 37.74\%, respectively. The corresponding Sharpe Ratios of 0.44, 0.68, and 1.11 show a negative correlation with $\lambda$, indicating that strong profitability compensates for the volatility brought by riskier investment choices.

Our Optimal Trading Technique produces positive returns under both bull and bear market trends. In the bull market of 2021, the OTT conducted 10,456 trades, while in the bear market of 2022, the OTT conducted only 3306 trades. Clearly, the quantity and quality of arbitrage opportunities are better in a bull market.

The applicability of the OTT is not limited to ETH. We extended our analysis to include popular cryptocurrencies BTC and SOL as intermediates in our pair trading strategy, as detailed in Appendix~\ref{sec:sensitivity}. This additional analysis validates the robustness and profitability of the OTT method across different cryptocurrencies and market conditions.

Despite the promising results, several limitations should be noted. Cryptocurrency carries significant risks for investors, as highlighted by several sources. For instance, \cite{swaminathan_anklesaria_aiyar_crypto_2022} likens the cryptocurrency market to the Dutch Tulip Bulb Bubble of the 17th century, where tulip bulb prices soared to extraordinary levels before collapsing dramatically. Additionally, many have documented the risks associated with cryptocurrency investments \citep{lee_cryptocurrency_2018, angerer_objective_2021}. We use cryptocurrency merely as an intermediate for pair trading fiat currencies due to its liquidity, 24/7 trading time, easy access to multiple fiat currencies in one market, and low barrier for data access. We do not promote investment in cryptocurrencies. Additionally, we selected two years of high-frequency data as our test data source, covering a full-cycle market, including a bull run and a bear run. However, this timeframe coincides with the COVID-19 period, during which fiat currencies were also volatile, contributing to spread movement. Profitability might differ in less volatile market conditions. Future work could extend the technique to model the spread from the perspective of the Ornstein–Uhlenbeck process to deepen the understanding of the market stochastic process.

\vspace{6pt}
\authorcontributions{Conceptualization, A.M.; methodology, H.Y.; software, H.Y.; validation, H.Y.; formal analysis, H.Y.; investigation, H.Y.; resources, A.M. and H.Y.; data curation, H.Y.; writing—original draft preparation, H.Y.; writing—review and editing, A.M.; visualization, H.Y.; supervision, A.M.; project administration, A.M.. 
 All authors have read and agreed to the published version of the manuscript.}

\funding{This research received no external funding.}

\institutionalreview{Not applicable.}

\informedconsent{Not applicable.}

\dataavailability{\label{sec:dataavail}The original data presented in the study are openly available from Kraken Exchange at Kraken Support 
 (\url{https://support.kraken.com/hc/en-us/articles/360047124832-Downloadable-historical-OHLCVT-Open-High-Low-Close-Volume-Trades-data}), accessed on 15 Apr 2024.}

\acknowledgments{The code implementing our methodology is available in a public repository:(\url{https://github.com/Hongshen-Yang/optimal-trading-technique}).}

\conflictsofinterest{The authors declare no conflicts of interest.} 


\appendixtitles{yes} 
\appendixstart
\appendix
\section[\appendixname~\thesection]{Sensitivity Analysis\label{sec:sensitivity}}
To further examine the robustness of our \acrfull{ott} method, we expanded the analysis to include additional cryptocurrencies and an extended time period. We included BTC (Bitcoin) and SOL (Solana) as intermediates for pair trading based on their market capitalization. Given that SOL is not directly tradeable against CAD, we focused on USD, GBP, and EUR as the common denominators. Additionally, since the selected full-cycle, bull-run, and bear-run markets coincide with the COVID-19 period\endnote{COVID-19, a pandemic since 2019, has had a significant global impact \citep{shi_overview_2020}.}, we extended the experiment to the post-COVID period using the latest publicly available data from Kraken up to 1 April 2024. Major cryptocurrencies have generally experienced an exponential growth, contributing to a strong Buy--Hold profitability.

\begin{table}[H]
  
  \caption{\label{tab:multi_corr} Correlation and cointegration among currency pairs.}
{
    \begin{tabularx}{\textwidth}{r CCCCCC}
      \toprule
       \textbf{Intervals} & \multicolumn{2}{c}{\textbf{15 min}} & \multicolumn{2}{c}{\textbf{60 min}} & \multicolumn{2}{c}{\textbf{720 min}} \\
      \midrule
      \textbf{BTC} & Corr & Coint & Corr & Coint & Corr & Coint \\
      \midrule
      USD:GBP & 0.9530 & (83.7\%) & 0.9504 & (87.5\%) & 0.9480 & (55.8\%) \\ 
      USD:EUR & 0.9687 & (76.0\%) & 0.9672 & (70.2\%) & 0.9659 & (31.7\%) \\ 
      GBP:EUR & 0.9806 & (94.2\%) & 0.9792 & (98.1\%) & 0.9776 & (73.1\%) \\
      \midrule
      \textbf{ETH} & Corr & Coint & Corr & Coint & Corr & Coint \\
      \midrule
      USD:GBP & 0.9773 & (69.2\%) & 0.9745 & (81.7\%) & 0.9717 & (79.8\%) \\ 
      USD:EUR & 0.9839 & (56.7\%) & 0.9822 & (70.2\%) & 0.9807 & (37.5\%) \\ 
      GBP:EUR & 0.9885 & (88.5\%) & 0.9873 & (95.2\%) & 0.9857 & (94.2\%) \\
      \midrule
      \textbf{SOL} & Corr & Coint & Corr & Coint & Corr & Coint \\
      \midrule
      USD:GBP & 0.9896 & (76.9\%) & 0.9892 & (86.5\%) & 0.9865 & (95.2\%) \\ 
      USD:EUR & 0.9932 & (60.6\%) & 0.9929 & (79.8\%) & 0.9918 & (84.6\%) \\ 
      GBP:EUR & 0.9958 & (89.4\%) & 0.9956 & (98.1\%) & 0.9936 & (99.0\%) \\
      \bottomrule
    \end{tabularx}}
\noindent{\footnotesize{This table provides average rolling correlation calculations and cointegration test results for each currency pair between 1 January 2022 and 1 January 2024.}}
\end{table}

Table~\ref{tab:multi_corr} demonstrates a relatively strong correlation among fiat currencies and a significant level of cointegration relationships. This period is characterized by a massive bull-run market, wherein the volatility among fiat currencies is negligible compared to the substantial appreciation of cryptocurrencies. Based on these observations, we will implement our \acrshort{ott} techniques on all three cryptocurrencies to assess their effectiveness across different market conditions.

\begin{table}[H]
  
  \caption{\label{tab:results_sens} Annualized profitability results for popular cryptocurrencies.}
\newcolumntype{R}{>{\raggedleft\arraybackslash}X}
      \begin{tabularx}{\textwidth}{r RRRRRR}
        \toprule
        & \multicolumn{2}{c}{\textbf{ETH}} & \multicolumn{2}{c}{\textbf{BTC}} & \multicolumn{2}{c}{\textbf{SOL}} \\
        \cmidrule{2-7}
        \textbf{Interval} & \textbf{BH} & \textbf{OTT} & \textbf{BH} & \textbf{OTT} & \textbf{BH} & \textbf{OTT} \\ 
        \midrule
        15 min (TC 0.1\%) & 357.88\% & 22.66\% & 246.99\% & 18.24\% & 1856.17\% & 7.42\%\\
        15 min (TC 0.0\%) & 358.80\% & 44.76\% & 247.68\% & 41.92\% & 1860.08\%& 68.42\%\\ 
        60 min (TC 0.1\%) & 357.88\% & 19.01\%& 246.99\% & 38.78\% & 1856.17\% & 3.70\%\\
        60 min (TC 0.0\%) & 358.80\% & 36.75\% & 247.68\% & 43.48\% & 1860.08\% & 33.67\%\\ 
        720 min (TC 0.1\%) & 357.88\% & 28.01\% & 246.99\% & 28.21\% & 1856.17\% & $-$0.80\%\\ 
        720 min (TC 0.0\%) & 358.80\% & 15.72\% & 247.68\% & 31.04\% & 1860.08\% & 3.26\%\\
        \bottomrule
    \end{tabularx}
\noindent{\footnotesize{The
arbitrage period is 1 October 2020 to 1 April 2024 (547 days). Participating assets are USD, GBP, and EUR. The abbreviations used in the table are BH (Buy--Hold), TC (Transaction Cost) and OTT (Optimal Trading Technique). Risk-adjust $\lambda$ is set to  1. \textit{open\_threshold} and \textit{close\_threshold} are tuned to the most lucrative combination.}}
\end{table}

The implementation of the \acrshort{ott} on ETH, BTC, and SOL traded under different time intervals is presented in Table~\ref{tab:results_sens}. Overall, the results demonstrated strong profitability not only for ETH but also for BTC. ETH and BTC, which dominate the cryptocurrency market, accounting for approximately 65\% of the total market cap \citep{coingecko_crypto_2024}, showed robust results in pair trading. While SOL achieved profitability in most trading scenarios, it did not match the performance of BTC and ETH. This is likely due to SOL's lower trading volume as it has, on average, less than two trades per minute against the GBP in the examined period and abnormal eighteen-fold annual growth. Despite the significant bull run experienced by major cryptocurrencies, our \acrshort{ott} method consistently produced positive returns, further validating its effectiveness across different market conditions. We measured the risk indicators of our implementation with $\lambda=1$ and transaction costs:
\begin{itemize}
    \item ETH has a standard deviation of 0.151 and a Sharpe Ratio of 1.24.
    \item BTC has a standard deviation of 0.159 and a Sharpe Ratio of 0.90.
    \item SOL has a standard deviation of 0.032 and a Sharpe Ratio of 1.07.
\end{itemize}

The volatility results display a similar property compared to the previously examined volatility in Table~\ref{tab:risknsharpe}, indicating that the technique maintained desirable market neutrality under a growth market. The Sharpe Ratio of ETH and BTC denotes the ability of OTT to generate stable yet profitable actions. Similar to profitability, SOL's low profit comes with a low standard deviation as well. With SOL's exponential growth in market value, the discrepancies in fiat currencies are negligible compared to the significantly volatile growth. We consider SOL as an exceptional case as it is extremely rare for a financial instrument to grow 44 times in just 18 months.

%

In conclusion, extending our experiment to include the post-COVID period provided additional validation for the robustness of our \acrshort{ott} method. Despite the unique market conditions during the pandemic, the method maintained its effectiveness, demonstrating its adaptability and resilience across different market environments. The application of multiple cryptocurrencies demonstrates the universality of the OTT, which is adoptable for not only the ETH but other financial instruments that meet the instrument assumptions. SOL, on the other hand, constitutes an outlier example of pair trading under extreme market conditions. This adaptability is crucial for traders seeking consistent returns amidst varying market dynamics.

\begin{adjustwidth}{-\extralength}{0cm}
\printendnotes[custom]
\reftitle{References}

\PublishersNote{}
\end{adjustwidth}
\end{document}